\documentclass[a4paper]{aipproc}
 
\layoutstyle{6x9}
  
\begin{document}

\title{A renormalisation-group approach to two-body scattering
with long-range forces} 

\author{Thomas Barford and Michael C. Birse}
{address={Department of Physics and Astronomy, University of Manchester,
Manchester, M13 9PL, U.K.}}

\begin{abstract}
We apply the renormalisation-group to two-body scattering by a combination 
of known long-range and unknown short-range forces. A crucial feature is that
the low-energy effective theory is regulated by applying a cut-off in the basis 
of distorted waves for the long range potential. We illustrate the method by 
applying it to scattering in the presence of a repulsive $1/r^2$ potential. We 
find a trivial fixed point, describing systems with weak short-range 
interactions, and a unstable fixed point. The expansion around the latter 
corresponds to a distorted-wave effective-range expansion.
\end{abstract}

\maketitle

\section{Introduction}

Effective field theories (EFT's) offer the promise of a systematic treatment
of few-nucleon systems at low-energies. (For a review, see: \cite{border}.)
They are based on the existence of a separation of scales between those of the
low-energy physics: momenta, energies, $m_\pi$ (generically $Q$), and those
of the underlying physics: $m_\rho$, $M_N$, $4\pi f_\pi$ (generically 
$\Lambda_0$). This makes it possible to systematically expand both the theory 
(defined by a Lagrangian or a potential) and physical observables in powers of 
$Q/\Lambda_0$. Such an expansion will be useful provided it converges rapidly
enough.

For weakly interacting systems (such as chiral perturbation theory in the zero- 
or one-nucleon sectors) the terms in the expansion can be organised according to 
naive (Weinberg) power counting. The order of a term $(Q/\Lambda_0)^d$ in the 
theory is just $d$ \cite{wein1,wein2}. 

In contrast for strongly interacting systems (such as $s$-wave nucleon-nucleon
scattering) there can be new low-energy scales generated by nonperturbative
dynamics. In such cases we need to resum certain terms in the theory and this 
leads to a new power counting \cite{bvk,vk3,ksw2,geg,bmr}, often referred to as
KSW power counting. The theoretical tool which allows us to determine the power 
counting in these cases is the renormalisation group (RG). 

\section{Renormalisation group}

The RG is an extension of simple dimensional analysis which has been used 
to study the scaling behaviour of systems in a wide range of areas of physics. 
In our work we use a Wilsonian version of the RG \cite{wrg}. This has the
following ingredients:
\begin{itemize}
\item{Impose a momentum cut-off, $|{\bf k}|<\Lambda$ on the low-energy 
effective theory.}
\item{Demand that observables be independent of the cut-off $\Lambda$. This
corresponds to integrating out physics on momentum scales above $\Lambda$.}
\item{Rescale the theory, expressing all dimensioned quantities in units of
$\Lambda$. Before doing this, we need to have identified all the important
low-energy scales in our system.}
\end{itemize}
We can then follow the flow of the various coupling constants of the theory
as $\Lambda\rightarrow 0$. If there is a clear separation of scales, then the 
theory should flow towards a fixed point. This is because, for 
$\Lambda<\!\!<\Lambda_0$, the only scale left is $\Lambda$ and so the rescaled 
theory becomes independent of scale.

Near a fixed point, perturbations scale as powers of the cut-off, $\Lambda^\nu$, 
and these define the power counting for the corresponding terms in the theory:
$d=\nu-1$. The sign of $\nu$ can be used to classify these terms:
\begin{itemize}
\item{$\nu>0$: irrelevant perturbation, flows towards the fixed point as 
$\Lambda\rightarrow 0$,}
\item{$\nu=0$: marginal perturbation (or ``renormalisable'' in field-theory 
terminology), leads to logarithmic flow with $\Lambda$,}
\item{$\nu<0$: relevant perturbation, flows away from the fixed point as 
$\Lambda\rightarrow 0$, making the point an unstable one.}
\end{itemize}

The application of these ideas to two-body scattering by short-range forces can 
be found in Ref.\ \cite{bmr}. Two fixed points were found: a trivial one 
describing a system with weak scattering, and a nontrivial one describing a
system with a bound state at zero energy. The expansion around the nontrivial 
one is organised by KSW power counting. It is in fact just the effective-range 
expansion \cite{ere}, reinvented in modern languange as an EFT.

\section{Distorted-wave theory}

Scattering by short-range interactions in the presence of a known long-range 
potential $V_L$ can be treated by distorted-wave theory. We write the full
scattering matrix as
\begin{equation}
T=T_L+(1+V_LG_L)\tilde T_S(1+G_LV_L),
\end{equation}
where $T_L$ is the scattering matrix for $V_L$ alone and $G_L$ is the 
corresponding Green's function.  The operator $\tilde T_S$ describes the
scattering between distorted waves of $V_L$. In terms of a short-range
potential $V_S$, it satisfies the Lippmann-Schwinger equation
\begin{equation}
\tilde T_S=V_S+V_SG_L\tilde T_S.
\end{equation}

We regulate this equation by cutting off $G_L$ in the basis of distorted 
waves \cite{kr,bb},
\begin{equation}
G_L=M\int^\Lambda{d^3{\bf q}\over (2\pi)^3}{|\psi_L(q)\rangle\langle\psi_L(q)|
\over p^2-q^2+i\epsilon}\;(+\hbox{bound states}).
\end{equation}
(Here $p=\sqrt{ME}$ denotes the on-shell relative momentum for two particles of 
mass $M$.) The use of this basis is crucial for the identification of the
scaling behaviour.

Since the long-range potentials of interest are generally singular as 
$r\rightarrow 0$, we cannot represent the short-range interaction by a 
simple $\delta$-function at the origin. Instead we choose a $\delta$-shell form,
\begin{equation}
V_S=V(p,\Lambda){\delta(r-R)\over 4\pi R^2}.
\end{equation}
Note that we have allowed the potential to be energy ($p$) dependent. However we 
have not included momentum dependence since, for the pure short-range-case, this 
was found to affect only the off-shell behaviour of scattering amplitudes 
\cite{bmr}.

We require that $V(p,\Lambda)$ vary with $\Lambda$ to keep the fully off-shell 
DW scattering matrix $\tilde T_S$ independent of cut-off. Rescaling the
resulting differential equation gives us a DW version of the RG 
equation \cite{bb}.

\section{Repulsive inverse-square potential}

To illustrate these ideas we apply them here to a specific example: a repulsive 
$1/r^2$ potential,
\begin{equation}
V_L={\beta\over r^2}.
\end{equation} 
This potential is analogous to the interaction in a three-body system
such as quartet $nd$ scattering, in the limit of an infinite two-body
scattering length \cite{efim}. It is scale independent and so
should be treated as part of a fixed point (and resummed to all orders). It acts 
in $s$-waves like a centrifugal barrier with ``angular momentum''
\begin{equation}
\lambda=\sqrt{\beta M+\textstyle{1\over 4}}-\textstyle{1\over 2}.
\end{equation}
The corresponding DW's are Bessel functions $j_\lambda(kr)$ of noninteger order.

The requirement that $\tilde T_S$ be independent of cut-off leads to a 
differential equation for $V(p,\Lambda)$,
\begin{equation}\label{vde}
{\partial V\over\partial\Lambda}=-{M V^2\over 2\pi^2}{\Lambda^2 
j_\lambda(\Lambda r)^2\over p^2-\Lambda^2}.
\end{equation}
Provided $R<\!\!<\Lambda^{-1}$ the DW factors scale as powers of $\Lambda$: 
$j_\lambda(\Lambda r)\propto (\Lambda R)^\lambda$. We can then rescale, defining
a dimensionless potential 
\begin{equation}
\hat V\propto MR^{2\lambda}\Lambda^{2\lambda+1}V,
\end{equation}
and on-shell momentum $\hat p=p/\Lambda$, to rewrite Eq.~(\ref{vde}) as an RG 
equation for $\hat V$.

We have found two-fixed points of this equation:
\begin{itemize}
\item{The trivial one, $\hat V=0$. Perturbations around this are of the form
$\Lambda^{2n+2\lambda+1}\hat p^{2n}$ and can be assigned an order
$d=2(n+\lambda)$.}
\item{A nontrivial one.  Perturbations around this are of the form
$\Lambda^{2n-2\lambda-1}\hat p^{2n}$ and can be assigned an order
$d=2(n-\lambda-1)$.}
\end{itemize}
All perturbations around the trivial point are irrelevant. In contrast,
the nontrivial fixed point is unstable; there is always at least one relevant 
perturbation.

\section{Discussion}

The method outlined here makes it possible to determine the power counting
for the short-range interactions in the presence of a known long-range potential.
A crucial feature is that the cut-off is applied in the basis of DW's. One
should not try to regulate the long-range as well as the short-range potential.

For the example described here, the repulsive $1/r^2$ potential, this leads 
(appropriately for Prague) to a baroque power counting involving non-integer 
orders. Nonetheless the terms in the EFT remain in one-to-one correspondance with
observables. For the expansion around the nontrivial fixed point these are the 
terms of a DW (or ``modified'') effective-range expansion \cite{ere,vhk}. If we 
write the full phase shift as $\delta=\delta_L+\tilde\delta_S$, where $\delta_L$ 
is due to $V_L$ alone, then this expansion has the form
\begin{equation}
p^{2\lambda+1}\left[\cot\tilde\delta_S-\cot\pi\left(\lambda
+\textstyle{1\over 2}\right)\right]=\sum_{n=0}^\infty C_{2n}\hat p^{2n},
\end{equation}
where $C_{2n}$ is the coefficient of the perturbation of order
$2(n-\lambda-1)$ in the potential.

We have also applied this method to the Coulomb potential, elucidating the
power counting for the results of Kong and Ravndal \cite{kr}, and
the attractive $1/r^2$ potential \cite{bb}. In the latter case, one has to
make the scattering for $V_L$ alone well-defined before the method can be
applied. This is done by choosing a self-adjoint extension \cite{pp} or, 
more physically, by including a short-range potential as part of $V_L$. 
The need for such an interaction has also been found in the corresponding 
three-body problems \cite{bhvk}.

\begin{theacknowledgments}
This work was supported by the EPSRC. MCB is grateful for the hospitality of
the Institute for Nuclear Theory, Seattle, where this work was started.
\end{theacknowledgments}

\end{document}